\documentclass{article}
\usepackage{latexsym}
\usepackage{graphicx}
\begin{document}

\title{Dynamical Systems Basis of Metamorphosis: Diversity and Plasticity
of Cellular States in Reaction Diffusion Network}

\author{Hiroaki Takagi\\
Laboratories for nanobiology, Graduate school of 
frontier biosciences\\
Osaka university, 2-2, Yamadaoka, Suita, Osaka, 565-0871, JAPAN\\
Kunihiko Kaneko\\
Department of Pure and Applied Sciences\\
University of Tokyo, Komaba, Meguro-ku, Tokyo 153, JAPAN}

\maketitle

\begin{abstract}
Dynamics maintaining diversity of cell types in a multi-cellular system
are studied in relationship with the plasticity of cellular states.
By adopting a simple theoretical framework for intra-cellular chemical reaction
dynamics with considering the division and death of cells, developmental 
process from a single cell is studied. 
Cell differentiation process is found to occur through instability in
transient dynamics and cell-cell interaction.  In a long time behavior, 
extinction of multiple cells is repeated, which
leads to itinerancy over successive quasi-stable multi-cellular 
states consisting of different types of cells.
By defining the plasticity of a cellular state, 
it is shown that the plasticity of cells decreases before the
large extinction, from which diversity and plasticity are recovered.
After this switching, decrease of plasticity again occurs, 
leading to the next extinction of multiple cells.  
This cycle of diversification and extinction is repeated.
Relevance of our results to the development and  evolution is
briefly discussed.
\end{abstract}

keywords: metamorphosis, diversity, plasticity, stability

\section{Introduction}
In multi-cellular organisms, developmental process from a single cell in 
embryo or a few homogeneous cells with multi-potency leads to an organism that 
consists of various cell types. 
Furthermore, in some multi-cellular organisms such as insects, 
the developmental process is generally accompanied by metamorphosis. 
There, after a cell society consisting of specific distribution
of some cell types is achieved and sustained over some time,
a transition to a novel society of distribution of different cell
types starts through specific process.  In other words,
there exists several multi-cellular states, each of which is
regarded to be quasi-stable distribution of cell types.
In the event of metamorphosis,
both the number of cells and the number of cell types decrease drastically, 
and then a different multi-cellular state is realized.
After this ``extinction'', adult organism makes embryos again to form the
next generation.  In other words, developmental process here
passes through a rather restricted state that is different from
the original embryo.
Through the process, adult body of each species is 
formed, from which the next generation is reproduced.  With this recursive
production,  the life cycle is repeated.

In usual developmental process, cells successively lose the multipotency,
that is ability to produce a set of different cell types.
This ability is regarded as a degree of changeability of a cell,
i.e., plasticity.
As the development proceeds, the number of cells and cell types
increases, while the plasticity of cells generally decreases.
Through the metamorphosis, however, some cells recover the ability 
to produce other type of cells, by regaining the plasticity.

In general, to consider the diversity of tissues and recursive production
of a multi-cellular organism, it is important to study how several stable 
distributions of different cell types are formed, sustained, or collapsed.
Then the following two general questions are addressed:\\
1. What mechanism causes the transition between different
quasi-stable states with several cell types?\\
2. How is the switching process related with the plasticity 
of cellular states?\\
Our purpose in the present paper is to answer these questions in terms 
of dynamics of diversity and plasticity of cell types.

As for the first question, ``isologous diversification theory'' was proposed 
\cite{Kaneko&Yomo2,Kaneko&Yomo3,Furusawa1,Furusawa2} as a theoretical 
framework for robust developmental process of multi-cellular organisms.
In these studies, dynamical systems approach is adopted, to show that 
robust developmental process with various cell types emerges 
as a result of interplay between intra-cellular dynamics and cell-to-cell
interactions. 
On the other hand, as is mentioned above, there are diversity at a tissue 
level, i.e., different stable distributions consisting
of different cell types, in a usual multi-cellular organism. So far, 
however, spontaneous formation of several quasi-stable 
states is not studied so much \footnote{In ref\cite{Furusawa2}, a preliminary
study for a proto-type of the life cycle of multi-cellular organisms is 
reported.}. 
Here we study the dynamics to maintain the diversity of cell types 
and the emergence of switches among several quasi-stable multi-cellular 
states. We show that all the cells change their states drastically at each 
switch, accompanied by deaths of multiple cells.

As for the second question, we show that the plasticity of
each cell decreases with the increase of diversity of cell types, through 
the developmental process. We discuss relationship between the two, and 
elucidate the switching mechanism in terms of the diversity and plasticity. 

For the present study, we first need to define plasticity of 
a cellular state.  Here we define it as changeability of a cellular state 
against external environmental change.  By choosing a specific model,
this plasticity is computed explicitly.  By studying a developmental
process of cells in the model, we will also find switches of multi-cellular
states accompanied by multiple cell deaths.  This switching
process will be shown to be tightly related with the loss of plasticity.
From extensive simulations, the long-term dynamics of cell society and
the plasticity we found are summarized as follows:\\
{\it As the development progresses, several cell types with low plasticity 
increase their number, which leads to extinction of several types of cells.
This extinction brings about a drastic change in environment surrounding cells.
Accordingly, internal states of all the surviving cells are changed.
Then, different cell types with high plasticity are generated.
From this `undifferentiated state', which is similar as the initial state, 
a new cell society with a different set of cell types is formed. 
In some case this cycle is repeated. }\\
We show that the cycle of increasing and decreasing diversity and
plasticity generally appears in our dynamical systems model. 
We reveal a general mechanism underlying commonly in such process, 
and formulate it in terms of dynamical relationship between diversity 
and plasticity of cell types.

For the purpose of the present study, we adopt a simple modeling framework
for internal chemical reaction dynamics, that is reaction-diffusion system 
on ``chemical species space'' . 
By adopting this modeling, one can correspond chemical concentrations with 
the fixed expression of genes, which is a common
representation for each differentiated cell type. We then discuss both the 
stability of realized cell states and their diversity.  

The present paper is organized as follows.
In the next section, we describe the details of our model.
We present the behavior of developmental process in section III.
The condition for the present model to show cell differentiation and 
switching over several quasi-stable cell societies is presented there.
In section IV, we reveal a feedback process leading to this switching, 
and confirm the relationship between the switching and the plasticity 
of total cells.
After presenting some results on a control of multi-cellular state by external
operations in section V, we briefly mention the generality of the
present result, and discuss its relevance to development and evolution in 
section VI.

\section{model}
In this paper,  we adopt a constructive modeling, by taking only some basic 
features of a problem in concern, to answer general questions. Here, with 
regards to diversity, stability and plasticity of cell states.

\begin{figure}[ht] 
\begin{center}
\includegraphics[width=5in,height=4in]{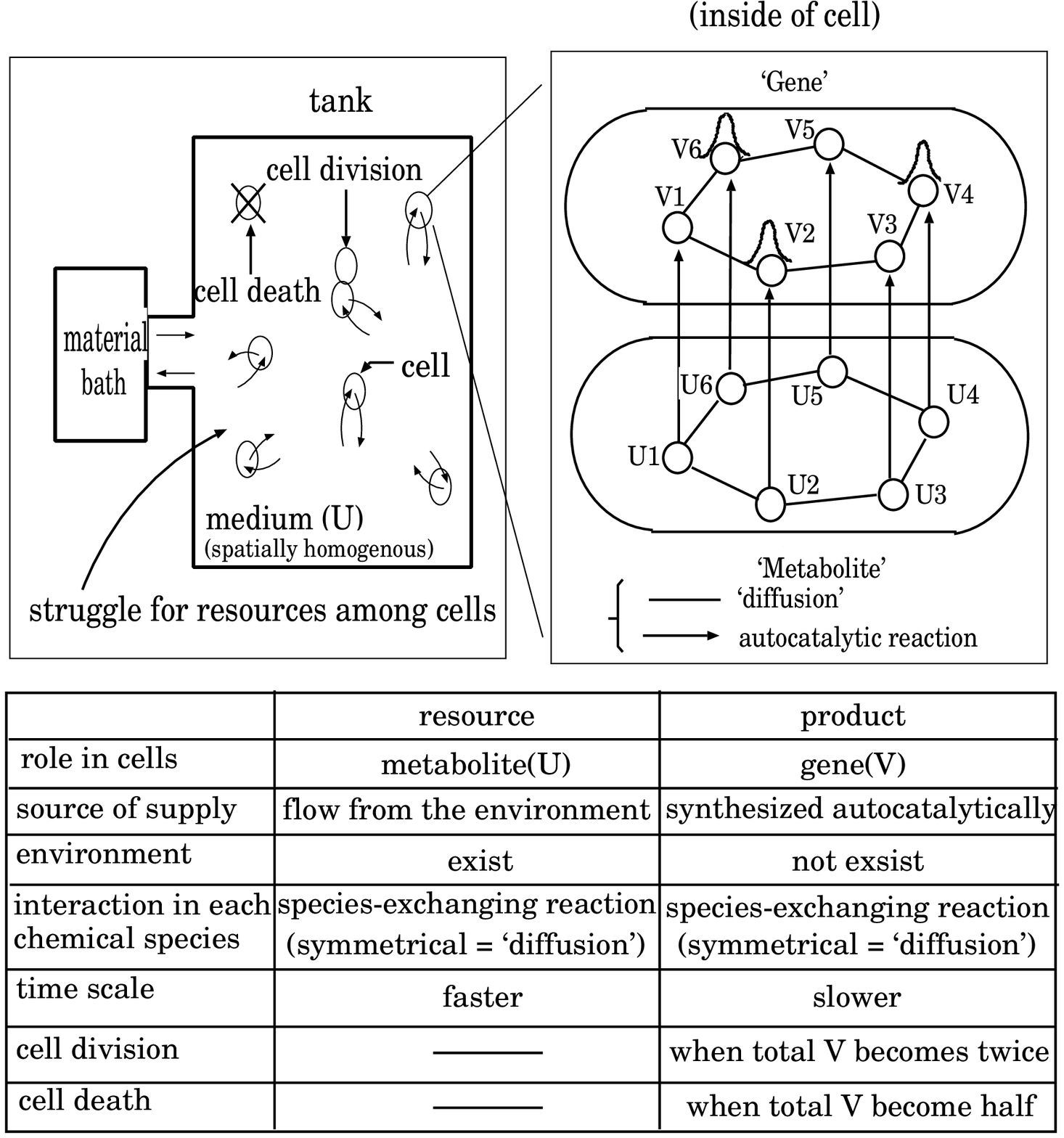}
\vskip 0.25cm
\caption{\small{Schematic representation of our model.}}
\label{fig1}
\end{center}
\end{figure}

The basic strategy of the modeling follows the previous works 
\cite{Kaneko&Yomo2,Kaneko&Yomo3,Furusawa1,Furusawa2}.
Cells with internal biochemical states compete for resources in the 
environment for their growth.
Following the growth or decrease of cell contents, a cell can divide or die 
so that the cell number changes in time.
Our dynamical systems model consists of the following three parts:
\begin{itemize}
\item intra-cellular chemical reaction network
\item cell-cell interaction
\item cell division and cell death
\end{itemize}
Now we describe each process. In Fig.\ref{fig1}, 
we show the schematic representation of our model.
\begin{itemize}
\item intra-cellular reaction network\\
In general, intra-cellular chemical reactions consist of
the reactions both of genes and metabolites.  Genes are set to
be on or off through the biochemical reactions.  This intra-cellular 
biochemical reaction constitutes a network both of genes and metabolites,
while the time scale for the reactions among genes are relatively much
slower than those of metabolites.
Here, simple intra-cellular chemical reaction dynamics is chosen as an 
abstract model, so that it satisfies basic features described above.

First, we assume a reaction network consisting of many 
product and substrate chemicals.  Existence of these two types of chemicals,
each constituting reaction networks, are inspired by the genes and 
metabolites in a cell.
The concentration of the $j$-th products in $i$-th cell
is denoted by $v_i^j(t)$, while that of substrates is denoted by $u_i^j(t)$.
Here, product chemicals are synthesized autocatalytically by consuming 
corresponding substrate chemicals. 
We adopt a variant of Gray-Scott model\cite{Gray&Scott,Pearson} as 
this autocatalytic reaction scheme\footnote{ Gray-Scott model is a 
reaction-diffusion system composed of two chemicals, substrate and product. 
It is a simplified version of autocatalytic Selkov 
model that explains self-sustained oscillation of glycolysis.}.
In this paper, we mainly present the results of the case where the number of 
chemical species is commonly set to $K=30$.

In the model $v_i^j(t)$ is assumed to correspond to the degree 
of expressions of a gene (or RNA), and $u_i^j(t)$ to concentration of 
a metabolite.
Besides the reaction dynamics between them, chemical concentrations may 
change through the reaction dynamics within a gene network or metabolic 
network. To take into account of such dynamics,
we assume reversible reactions in each of product and
substrate chemicals, which form two reaction networks. 
Each chemical is converted to other chemicals by a reversible reaction 
given by the network. 
The reaction network is represented by a reaction matrix $W{(i,j)}$, 
which is 1 if there is a reaction from chemical $i$ to 
chemical $j$, and 0 otherwise. Since the reaction network is assumed to be 
reversible, the matrix is symmetric, i.e., if $W{(i,j)}=1$, then $W{(j,i)}=1$. 
Here, we adopt the same network for products and substrates, and also assume
that the network is composed of two reaction paths per chemical to form a 
single closed-loop structure for simplicity.

The rate constant of reversible reaction $C_{u}$ and $C_{v}$ are
assumed to be common to all resources and all products, respectively.
Moreover, we also assume that $C_{u}$ is larger than $C_{v}$, by considering
the difference in the time scales between metabolites and genes. 
These values are fixed throughout the simulation.
Then, these reversible reactions are regarded as `diffusion' on 
`chemical species space', so that intra-cellular reaction between products 
and substrates is regarded as a one-dimensional reaction-diffusion system\cite{Turing}. 
In that representation, each attractor of intra-cellular reaction dynamics
corresponds to a cell state with a different genetic expression pattern. 
Since developmental process in real cell system
is elucidated as spontaneous cascading processes with different gene 
expression patterns, we study how different cellular states
are selected by including the developmental process to the model.
 
\item cell-cell interaction\\
Assuming that environmental medium is completely stirred, we can neglect 
spatial variation of chemical concentrations in it, so that all the cells
share the spatially homogeneous environment.
Here we consider only the diffusion of resource chemicals 
through the medium as a minimal form of interaction.
In this model, we assume that only substrate chemicals are transported through 
the membrane as resources, in proportion to the concentration difference 
between the inside and the outside of a cell. All the resource chemicals have 
the same diffusion coefficient $D_{u}$. 
Each cell grows by taking resource chemicals from the medium and transforms 
them to product chemicals.
$U^j$ is the concentration of $j$-th resources in the medium.
Resource chemicals in the medium are consumed by cells, while we assume 
that resource chemicals are supplied from external material 
to the medium, with the rate proportional to the difference between the 
concentration in the bath and the medium. 
Again, all the resource chemicals have the same diffusion coefficient 
$D_{U}$ in the medium. 
The concentrations of all resources in the material bath, $U$ are
set to be 1. The parameter $Vol_0$ is the volume ratio of a medium to a cell,
and $N$ is the number of cells.

\item cell division and cell death\\
Each cell gets resource chemicals from the medium and grows by 
transforming them to the product chemicals.
Here we assume that cell volume (denoted by $Vol_i$ for cell i) is 
proportional to total amount of product chemicals. 
For each time step in temporal evolution, we count the change of 
product chemicals, that is, the change of total cell volume, and each 
chemical concentration becomes to be normalized by the factor. 
When the cell volume becomes twice the original, then the
cell is assumed to divide, while if it is less than half the original,
then the cell is put to death.
In real biological system, cell division occurs after replication of
DNA (which has smaller diffusion coefficient and cannot penetrate through
the membrane). Hence these assumptions are rather natural.  
After cell division, each cell volume is set to be half, and
each cell is divided into two almost equal cells, with some fluctuations. 
To be concrete, chemical concentration $a$ ($a$ is a representation of 
$u$, $v$)
is divided into $(1+\eta)a_i^{(j)}$ and $(1-\eta)a_i^{(j)}$ respectively,
where $\eta$ is a uniform random number over $[-10^{-3},10^{-3}]$.
These fluctuations can give rise to a small variation among cell states, 
which eventually leads to cell differentiation by being amplified through 
cell-cell interaction.
\end{itemize}
Accordingly, the concentration change of each chemical species is given by
\begin{eqnarray}
\Delta u_i^j(t)=D_{u}(U^j(t)-u_i^j(t))-u_i^j(t)v_i^j(t)^{2}\nonumber\\
\Delta v_i^j(t)=u_i^j(t)v_i^j(t)^{2}-Bv_i^j(t)
\end{eqnarray}
\begin{eqnarray}
du_i^j(t)/dt=\Delta u_i^j(t)+C_u\sum_{k=1}^{K}W{(j,k)}(u_i^k(t)-u_i^j(t))-u_i^j(t)dVol_i(t)/dt\nonumber\\
dv_i^j(t)/dt=\Delta v_i^j(t)+C_v\sum_{k=1}^{K}W{(j,k)}(v_i^k(t)-v_i^j(t))-v_i^j(t)dVol_i(t)/dt\nonumber\\
dU^j(t)/dt=\frac{D_{u}}{Vol_0}\sum_{i=1}^{N}(u_i^j(t)-U^j(t))+D_{U}(U-U^j(t))\nonumber\\
dVol_i(t)/dt=\frac{\sum_{j=1}^{K}\Delta v_i^j(t)}{\sum_{j=1}^{K}v_i^j(t)}Vol_i(t)\nonumber
\end{eqnarray}
Here, the last term for each equation for 
$du_i^j(t)/dt$ and $dv_i^j(t)/dt$ represents the dilution
of concentration by the increase of the volume.

In the model introduced above, a single cell has many 
fixed-point attractors, in contrast to the previous studies \cite{Kaneko&Yomo2,Kaneko&Yomo3,Furusawa1,Furusawa2}, where a single cell can take only one or 
a few attractors.
Many stable cellular states are realized accordingly, that correspond to
different cell types. Furthermore, as will be shown, by cell-cell interaction,
cells are differentiated to take different chemical compositions.

\section{Developmental process with cell division and cell death}
The behavior of a single cell state depends on the bifurcation structure
of the reaction-diffusion system on `chemical species space' mentioned above.
In the region where uniform steady state becomes unstable, a single 
cell state has multiple attractors. The dependence of the number of attractors
on the number of chemical species is expected to be exponential, which is
verified in numerical simulations(data not shown).  
Based on these results, we discuss developmental process with the change of 
cellular states under the process of cell division and cell death. 
We study a coupled dynamical system, where the intra-cellular state and 
the inter-cellular interaction are mutually influenced. 
By cell-cell interaction, a homogeneous cell society of a single cell 
attractor may be destabilized, so that novel states may appear. 
We study such cell differentiation processes here.

\subsection{Initial conditions and methods}
In all the simulations, we mainly set the parameters $C_u=2.0$, 
$C_v=0.020$, $D_{u}=0.50$, $D_{U}=1.0$, $Vol_0=3.0$, $B=0.060$. These values 
of $B\sim C_v$ correspond to one of typical values at which the original 
Gray-Scott model in one-dimensional space forms a self-replicating spot 
pattern, so that our model also shows such pattern dynamics.
Initial condition of the first cell is chosen as $u_i^j(0)=0.50$ and 
$v_i^j(0)=0.250+0.01\times rand(j)$, where $j=1,2,...,K$ 
and $rand(j)$ is a uniform random variable over [-1,1], although this 
specific choice is not important.

\subsection{Cell differentiation}

\begin{figure}[ht]
\begin{center}
\includegraphics[width=7in,height=7in]{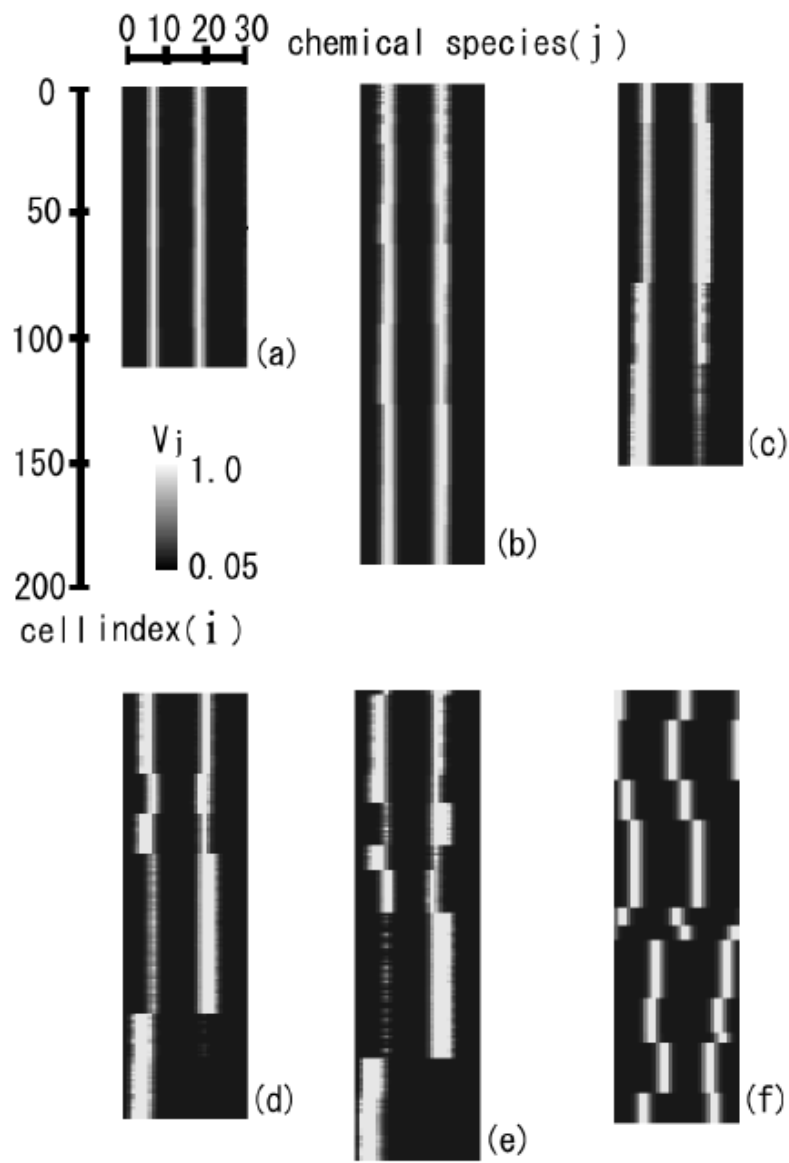}
\caption{
An example of the time series of all the cells. Snapshot patterns of the 
concentrations of all the product chemicals are overlaid with a gray scale
at every $t=50$((a)$\sim$(e)) except for (f)(at $t=10000$).
The horizontal axis represents the number of product chemicals, whereas 
the vertical axis represents the indices of cells, which are sorted so that 
the cells of the same type are aligned.
Unless otherwise mentioned, we adopt the parameter values $A=0.020$,$B=0.060$,
$C_u=2.0$, $C_v=0.020$, $D_{u}=0.50$, $D_{U}=1.0$, $Vol_0=3.0$ and $K=30$ for 
later figures.
}
\label{fig7}
\end{center}
\end{figure}

Now we show an example of the differentiation process. 
Fig.\ref{fig7} shows a typical temporal evolution of the concentration of $v^j (j=1,..,K)$ for all the cells starting from a single cell, in which cell 
differentiation occurs. A series of snapshots are shown by using a gray scale 
for the concentration of product chemicals. Each group of distinct cellular 
states with a different set of values $v_i^j$ corresponds to a quasi-stable 
cell type, which makes recursive production. 
(Time for reproduction is much longer than typical transient time to reach 
each quasi-stable state.) 
The figure also includes the cell states that appear at a later stage of
the temporal evolution(Fig.\ref{fig7}(f)). There exist eleven types, 
all of which are fixed points.
Under the instability of transient state of cells, small fluctuations in cell 
division process are amplified through the competition for resources in 
environment among all the cells, which leads to cell differentiation. 
From extensive simulations of the present model,
the necessary conditions for cell differentiation are summarized as follows
\footnote{If we set our initial condition of the first cell exactly on an 
attractor of a single cellular state, differentiation does not occur.
In this case, cells will go extinct after several divisions.}:\\
(1) Inter-cellular interaction is stronger than some threshold.\\
(2) The ratio of the reaction rates for $u^j$ to that for $v^j$, i.e.,
$C_{u}/C_{v}$, is in the intermediate range.\\
(3) The number of chemicals is larger than some threshold.\\
Details on these conditions are shown in appendix.

\subsection{Switching between quasi-stable ensembles consisting of several cell types}
\begin{figure}[ht]
\begin{center}
\includegraphics[width=5in,height=3in]{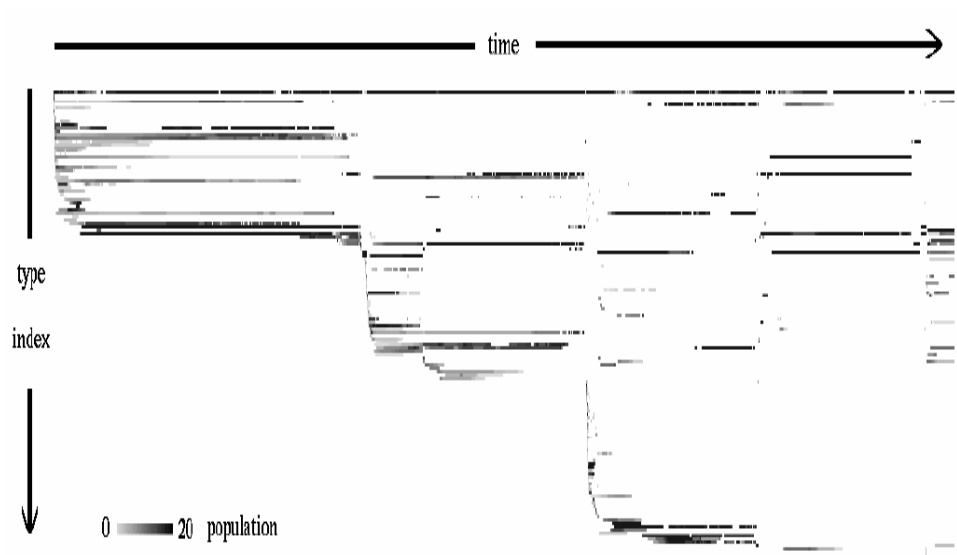}
\vskip 0.25cm
\caption{
\small{An example of the long time behavior of cell differentiation process.
Simulation is carried out up to $t=1000000$ starting from a single cell, 
while the data for the total cells are sampled by every $1000$ time, to 
classify all cell types and to get the temporal change of population of 
each cell type. All cell types are shown in the order of their appearance;
a cell type that appears earlier in the simulation has a smaller
index for the cell type. The population of each cell type is represented 
with a gray scale. Here the unit time of the figures is 1000.
}}
\label{fig9}
\end{center}
\end{figure}

Next we investigate the long time behavior of cell differentiation process.
We show an example of typical temporal evolution by plotting 
existing cell types and their population at every 1000 time unit in 
Fig.\ref{fig9}. A multi-cellular state consisting of several cell types 
is formed at a very early stage,
and is maintained over a long period, which is much longer than typical 
transient time of a single cell state. Then a sudden crisis occurs, and 
after some generations, a new multi-cellular state with 
different cell types is recovered.
This crisis is accompanied by the decrease of the cell number and 
diversity in cell types.
This switching between diversification and extinction can occur repeatedly.

\section{Further Analysis of the switching process}
\subsection{quantitative characterizations of the plasticity of cell types}
Now we investigate the switching among several quasi-stable multi-cellular 
states in more detail. 
We define recursiveness of each cellular state through the reproduction, 
by comparing the average concentrations of product chemicals between a
mother cell and its daughter cell.  
This similarity of chemical compositions between a mother and its daughter 
cell is defined as follows.
Let $\overrightarrow{\rm V_i(n)}$ denote the vector representation of 
all the average concentration of product chemicals ($v_i^j$) of the i-th 
cell between (n-1)-th and n-th cell division, that are denoted by 
$\overline{v_i^j}$ for $j=1,2...,K$.
As an index for the recursiveness of cellular state, we introduce an inner 
product denoted by $H_i(n)$ between $\overrightarrow{\rm V_i(n)}$
and $\overrightarrow{\rm V_i(n-1)}$.
\begin{center}
$H_i(n)=\frac{\overrightarrow{\rm V_i(n)} \cdot \overrightarrow{\rm V_i(n-1)}}{|\overrightarrow{\rm V_i(n)}||\overrightarrow{\rm V_i(n-1)}|}$,
$\overrightarrow{\rm V_i(n)}=(\overline{v_i^1},\overline{v_i^2},...,\overline{v_i^{K}})$
\end{center}
If $H_i(n)\approx1$, the cell is regarded to keep the same type 
between successive cell divisions, that means recursive reproduction.
In the present model, each differentiated cell type is represented as 
distinctly separated chemical states of cells where each cell type can 
produce its own type recursively. 
The cell differentiation processes in our model are well represented with 
these two quantities, by which we can distinguish all cell states correctly.

Next we introduce a measure of plasticity of cell type to study the switching 
process in terms of the temporal change of plasticity of cell types.
Here we define the plasticity of each cell type as changeability 
of the cell state against environmental fluctuations, which are the change 
of concentrations of resource chemicals in the environment caused by cell 
divisions and deaths. 
We characterize the plasticity of cell state in the following manner:\\
First we take ${v{(o)}_i}^j$, the concentration of $j$-th product chemical 
of the type $i$ cell, while we define ${v{(f)}_i}^j$ as the concentration 
of $j$-th product chemical of the attractor of the cell type $i$ by 
eliminating cell division and death processes, to check the attractor 
state at the fixed environment. 
Then we define the {\bf ``attractor distance'' $Dist_i$} of the type $i$ cell, 
as the Euclid distance between ${v{(o)}_i}^j$ and ${v{(f)}_i}^j$, namely,
$Dist_i=\sqrt{\sum_{j=1}^{K}{({v{(f)}_i}^j-{v{(o)}_i}^j)^{2}}}\nonumber$.\\
As the distance is smaller, the cell type is closer to a specific attractor
reached under a fixed environment by fixing cell-cell interactions.

Now, we conjecture that 

(1)the cell plasticity (changeability) is characterized
by this attractor distance (i.e., distance from an attractor)
and that 

(2) the potentiality of differentiation  
decreases as this attractor distance is smaller.

We will demonstrate these conjectures by measuring the temporal flucutuations
of cells and frequency of differentiation.

{\bf conjecture (1)}:
First, we note that the plasticity of a cell means the changeability
of the state against the environmental change.  
Here, the conjecture means that this changeability
is smaller as the state is closer to a given attractor.
In the present system, the environment fluctuates accoding to the birth 
and death of surrounding cells. Hence, the changeability against environment
is computed by the temporal fluctuations
of chemical compositions of each cell type.
(Such relationship between flucturation and response agaisnt environment
is known as fluctuation-disspiation theorem in physics, and is extened
to a biological system recently\cite{Sato}.)

\begin{figure}[ht]
\begin{center}
\includegraphics[width=4in,height=2in]{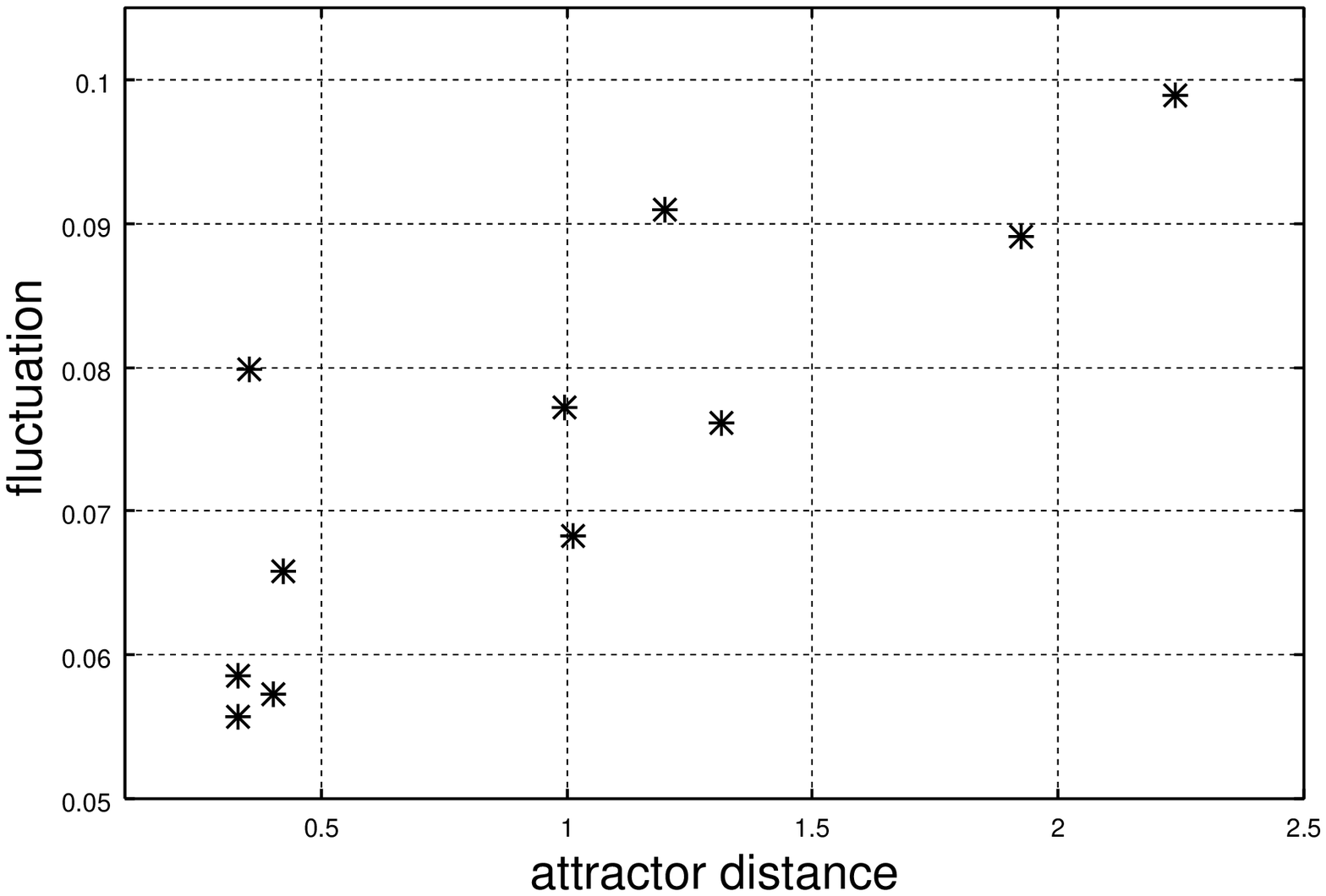}
\caption{
\small{
An example showing the relation between the attractor distance and the 
temporal fluctuation of each cell type. 
Over a period when several types coexist stably in a temporal evolution, 
we measured the temporal fluctuation of each cell type as 
the Euclid distances of chemical concentrations between two cells
of the same type, chosen at different 
time, separated by a time span $t=1000$.  The
temporal fluctuations are computed over $100$ data samples, and
the average over the samples are plotted.  
}}

\label{fig10}
\vskip 0.25cm
\includegraphics[width=4in,height=2in]{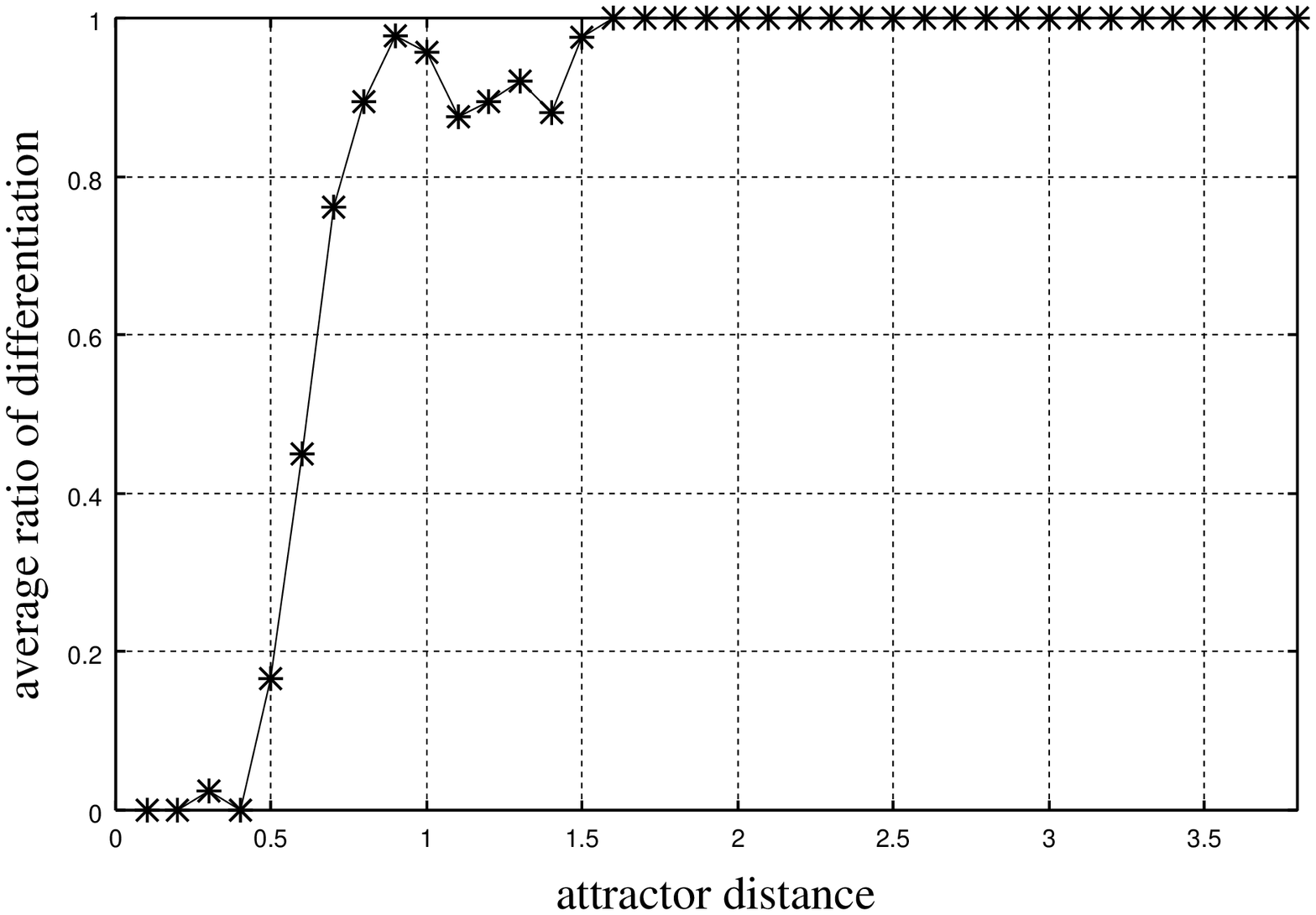}
\caption{
\small{The relation between the attractor distance and the frequency of 
re-differentiation. We compute attractor distance of all the cell types 
appeared in each temporal evolution, and take an ensemble of cells whose 
members have the same initial condition.  This cell ensemble
is put into a new environment to check
whether cell differentiation occurs or not.
By sampling the data for the attractor distance by 0.1 bin size,
we compute the average ratio of the frequency of re-differentiation event 
for each bin, to get the relationship between the differentiation ratio 
and the attractor distance.
}}
\label{fig11}
\end{center}
\end{figure}

Hence, we demonstrate the conjecture (1) by
checking if there is positive
correlation between the attractor distance and the degree of temporal 
fluctuations of each cell state.
We compute the temporal fluctuation of each 
cell type for some periods where several cell types coexist stably in each 
temporal evolution. 
This temporal fluctuation of each cell type is computed as the  
difference of chemical concentrations (i.e., Euclid distance)
between two cells at a given time span.
Here we take the time span $t=1000$, and temporal 
fluctuations are calculated from the average over $100$ data samples.  
An example of this temporal fluctuation plotted as a function of
the attractor distance is shown in Fig.\ref{fig10}, which shows
positive correlation between the two.
We investigate other data over ten periods from five temporal 
evolutions including an example of Fig.\ref{fig9}, and all of them show 
positive correlations. (The correlation coefficient
ranges from 0.46 to 0.84.)
This result shows positive correlation between
the temporal fluctuations of the chemical concentrations of a cell type 
and its attractor distance. 

{\bf Conjecture (2)}:
The second conjecture
will be confirmed by the positive correlation between 
the attractor distance and the frequency
of the occurrence of re-differentiation event when a cell enesmble
consisting of cells of this single cell type
are put to a new environment.
We compute the average of this frequency for each cell type, over
cells within a given range of attractor distance.  The relationship
between the frequency of re-differentiation and the attractor, thus
obtained, is plotted in Fig.\ref{fig11}, using the same data as above.
\footnote{Here the population of each cell type is set to be 80. 
If the number is much larger, the homogeneous cell ensemble grown from 
the group goes extinct by the lack of the resources. On the other hand, 
if the number is much smaller, then the effect to internal cell states 
caused by the change of the environment is so large that the cells lose 
the original character when they are selected.
Hence we choose this medium number for cell population.}
As shown in Fig.\ref{fig11}, there is a sigmoidal function dependence, 
between the attractor distance and the frequency of re-differentiation.
There is a threshold for the attractor distance below which
the corresponding cell type loses the ability of re-differentiation 
drastically. 
\footnote{This also supports the initial condition dependence 
of cell differentiation event mentioned previously.}

By summing up these two results, it is confirmed that the attractor distance 
introduced above is valid as a measure of plasticity of cell type.

\subsection{Mechanism for switching through extinction of many cells}

Now we discuss the switching with multiple cell deaths in
relationship with the loss of plasticity.

First, we summarize our results discussed here as the following scenario:
At each stage of given quasi-stable multicellular states,
cell types with different degrees of plasticity coexist.
Then at each stage, cell types with relatively high plasticity (i.e., with
larger attractor distance) differentiate to other cell types with lower 
plasticity, 
so that the ratio of cell types with lower plasticity increase gradually.
Then the distribution of cell types allowing for effective use of resource 
chemicals is destroyed, resulting in extinction of many cell types.
With these multiple deaths,  the concentrations of environmental resource 
chemicals change drastically, leading to re-differentiation of some of
surviving cells with low plasticity into a new state with high 
plasticity. Then, emergence of novel cell types with high plasticity 
gives rise to a novel 
multi-cellular state, and effective use of resources becomes possible again.
With this drastic change, switch to a novel multi-cellular state follows.

\begin{figure}[ht]
\begin{center}
\includegraphics[width=4in,height=2in]{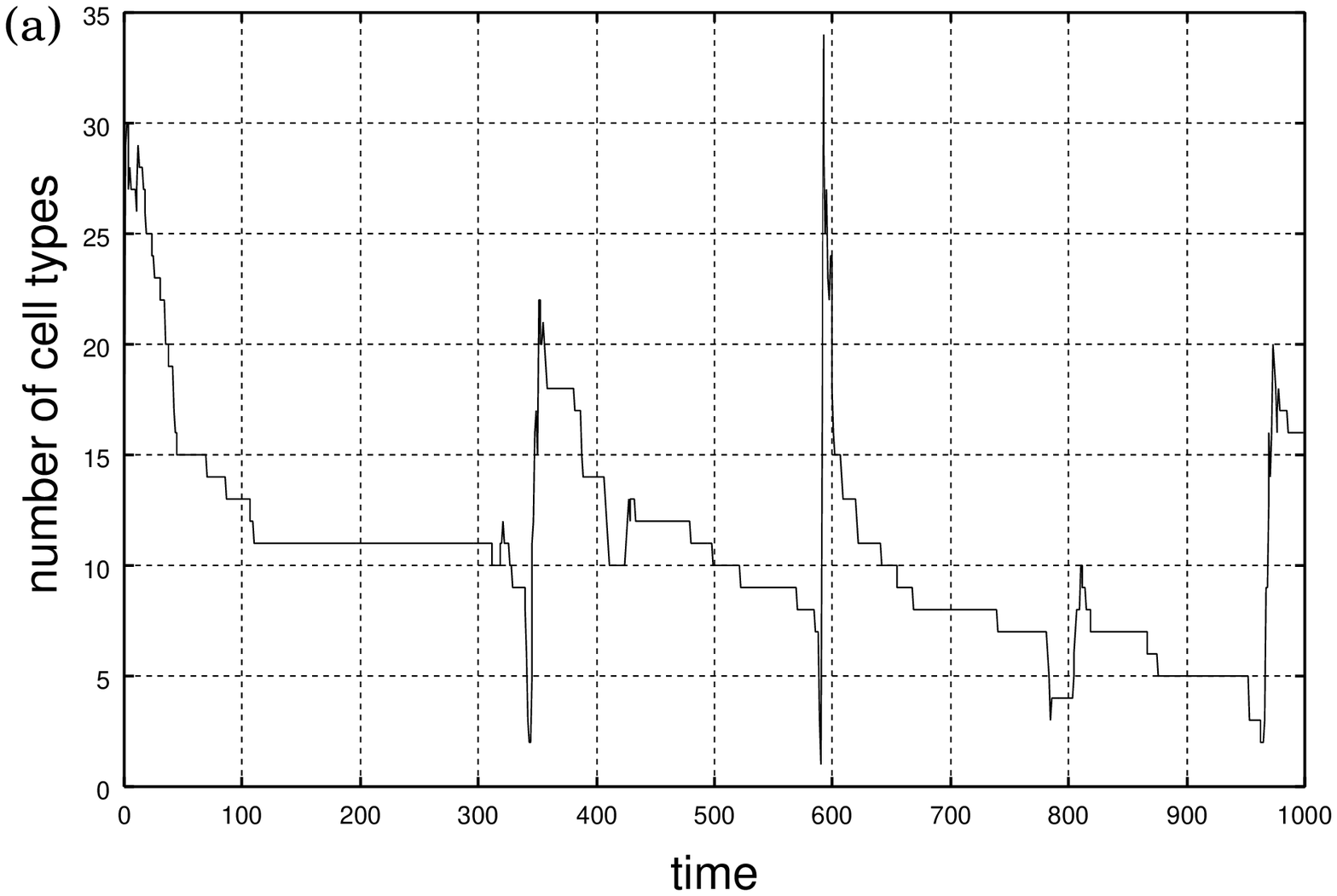}
\includegraphics[width=4in,height=2in]{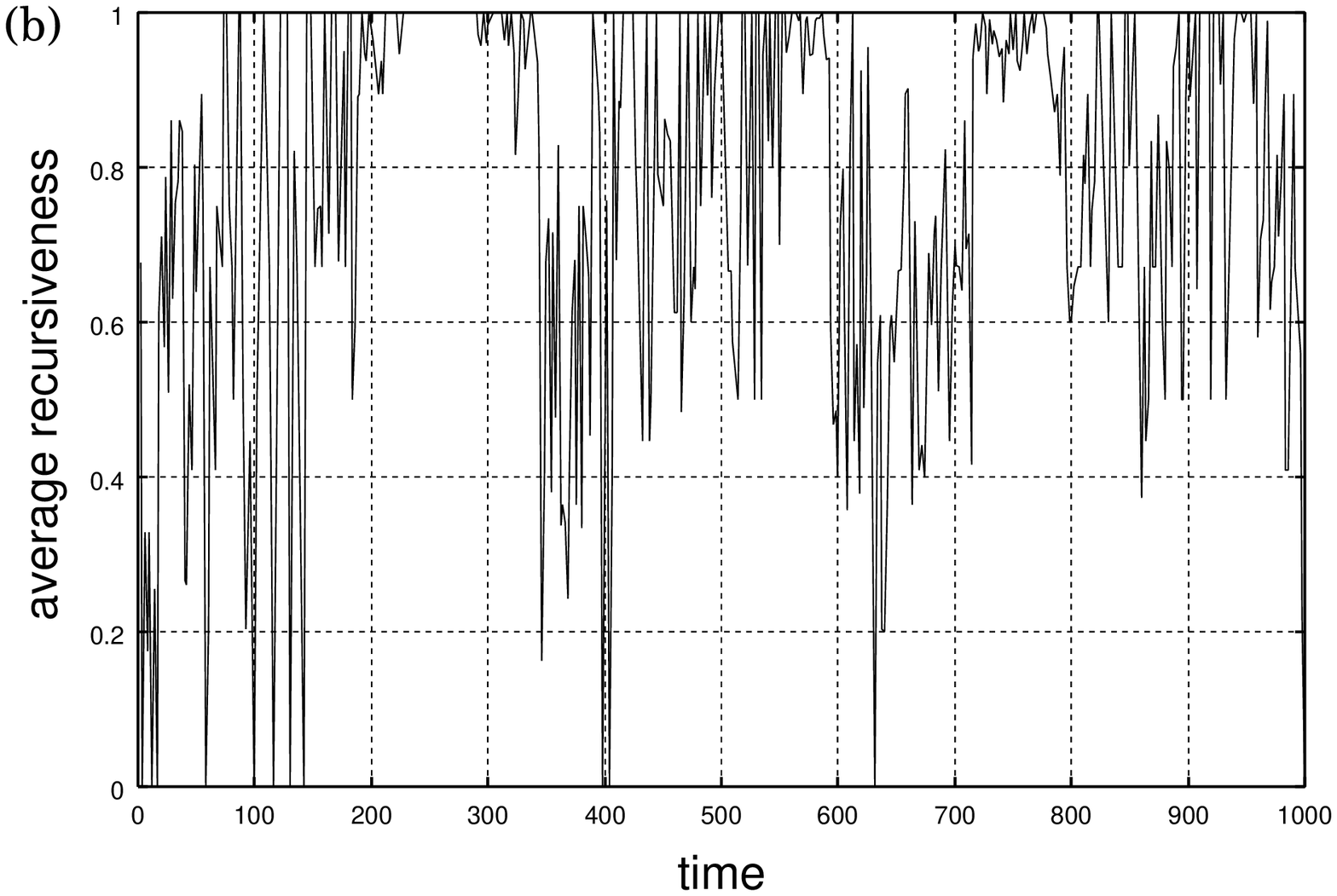}
\includegraphics[width=4in,height=2in]{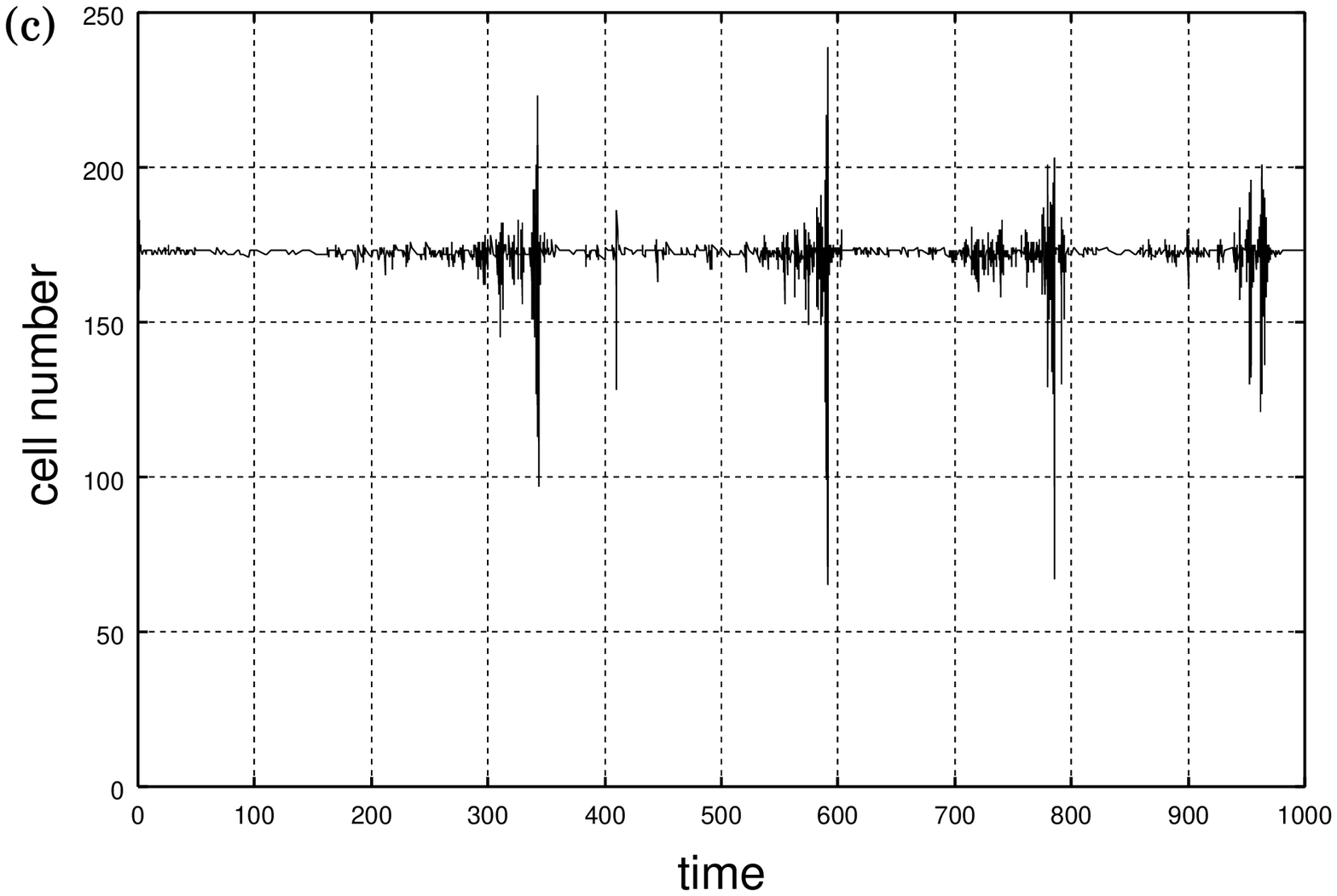}
\vskip 0.25cm
\caption{
\small{(a):Temporal change of the number of cell types.
(b):Temporal change of the average recursiveness over all cells.
(c):Temporal change of the total cell number.
We computed all the quantities for the data given
in the temporal evolution of Fig.\ref{fig9}.
Here the unit time of the figures is 1000.}}
\label{fig12}
\end{center}
\end{figure}

\begin{figure}[ht]
\begin{center}
\includegraphics[width=4in,height=2in]{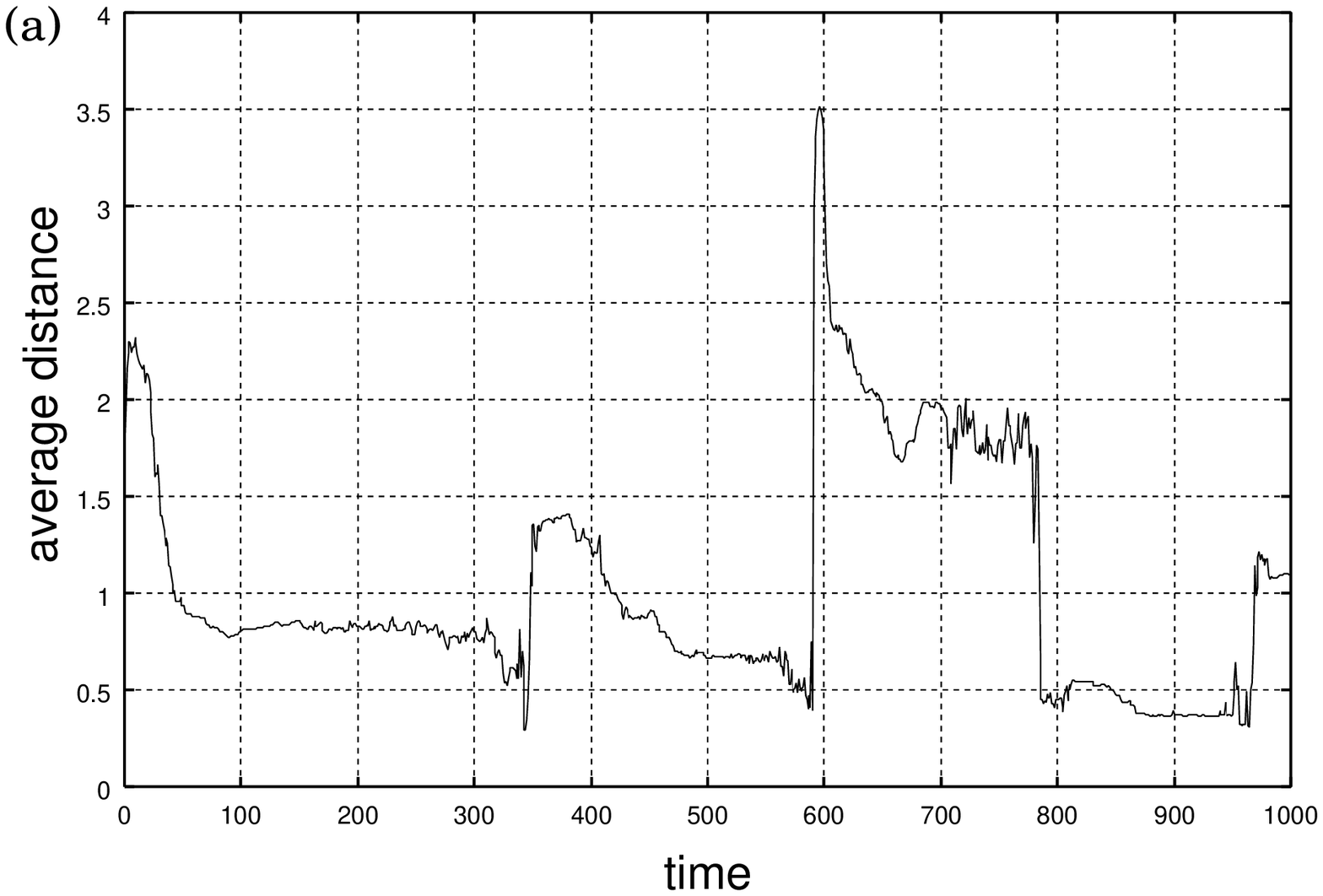}
\includegraphics[width=4in,height=2in]{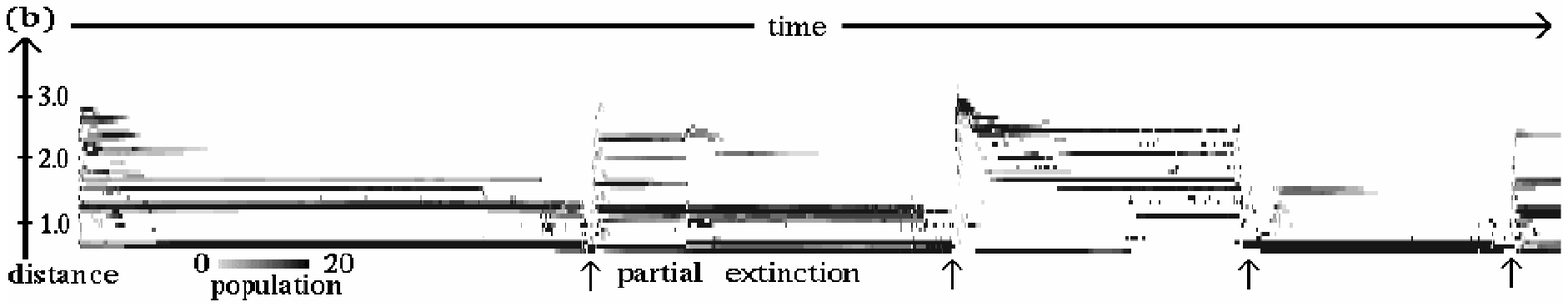}
\vskip 0.25cm
\caption{
\small{(a):Temporal change of the average attractor distance.
The average attractor distance is obtained by averaging over all existing
cells. 
(b):Temporal change of attractor distances of all the cell types existing
simultaneously are plotted with their population.
We computed all the quantities for the data given
in the temporal evolution of Fig.\ref{fig9}.
Each time when extinctions of many cells occur is shown by the arrow.
Here the unit time is 1000.}}
\label{fig13}
\end{center}
\end{figure}

This scenario is verified by computing the temporal change of 
the following five quantities using the data for the 
temporal evolution of Fig.\ref{fig9}:
the total diversity of cell types (Fig.\ref{fig12}(a)), the average 
recursiveness of cell types over all cells (Fig.\ref{fig12}(b)), the total 
number of cells (Fig.\ref{fig12}(c)), the average of attractor distances 
over all of the existing cell types at each moment(Fig.\ref{fig13}(a)) 
and the number of cells at each bin of attractor distances(Fig.\ref{fig13}(b)).
As shown in Fig.\ref{fig12} and Fig.\ref{fig13}, the average attractor 
distance as well as the number of cell types decreases first.
With these decreases, the recursiveness of cells increases on the average.
Then, the attractor distance and the number of cell types
stays at low values, and almost complete recursive production is 
sustained, since the most existing cells have low plasticity. 
After slight decrease of the 
attractor distance and diversity, then, these two values
go up to higher values, 
accompanied by multiple cell deaths and switching of cell types.  Now,
high plasticity and diversity of cell types are recovered.
After this recovery, the attractor distance and diversity
again decrease gradually, until the next multiple cell deaths occur.
This cycle consisting of the decrease of diversity, extinction, and 
recovery is repeated.

Although we show only one example here, qualitatively 
the same behavior is generally observed at each switching event
in the present model. 
The scenario mentioned above is rather universal.

\section{Transition of states by external operations}

So far, we have studied the switching process with regards to relationship 
between diversity and plasticity of cell types.
In this section, we study the behavior of the present cell system
after some kind of operation is applied to the cell system.
First we add noise into intra-cellular chemical reactions.
Here the noise is regarded as the fluctuation of the number of molecules,
and is assumed to be Gaussian white noise. 
The stochastic differential equations for resource and product chemicals 
in a cell are expressed by adding a noise term
to the ordinary difference equations (1);\\
$\Delta v_i^j(t)=u_i^j(t)v_i^j(t)^{2}-Bv_i^j(t)+\eta\sqrt{v_i^j(t)}$\\
$\Delta v_i^j(t)=D_{u}(U^j(t)-u_i^j(t))-u_i^j(t)v_i^j(t)^{2}+\eta\sqrt{u_i^j(t)}$.\\
Here $\eta$ is a Gaussian white noise satisfying $\langle\eta(t)\eta(t')\rangle=\sigma^2\delta(t-t')$.

\begin{figure}[ht]
\begin{center}
\includegraphics[width=4in,height=2.5in]{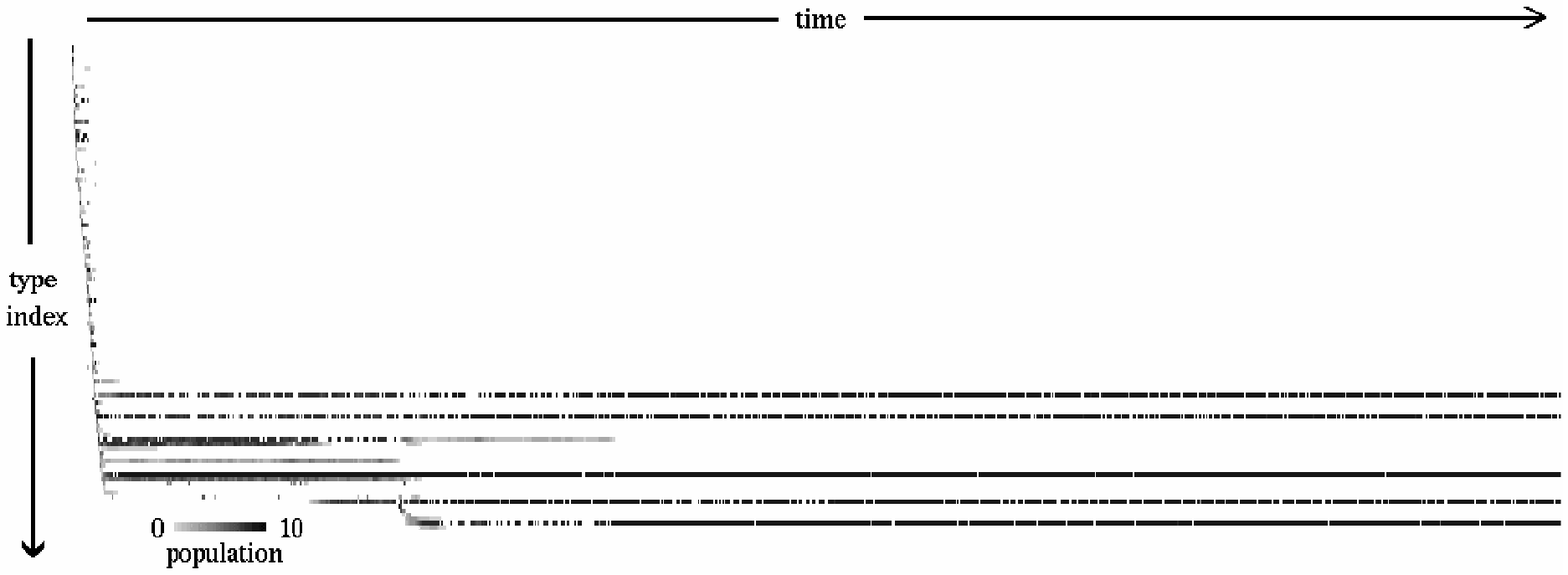}
\caption{
\small{An example of the long time behavior of cell differentiation process
by adding noise with the amplitude 0.00010.
Simulation is carried out up to $t=1000000$ starting from a single cell, 
while the data for the total cells are sampled by every $1000$ time, 
to classify all cell types. 
The initial condition and method for the classification of cell types are 
mentioned in the text.
All cell types are shown in the order of their appearance;
a cell type that appears later in the simulation has a larger
index for the cell type.  The
population of each cell type is represented with a gray scale.
Here the unit time is 1000.}
}
\label{fig14}
\includegraphics[width=4in,height=2in]{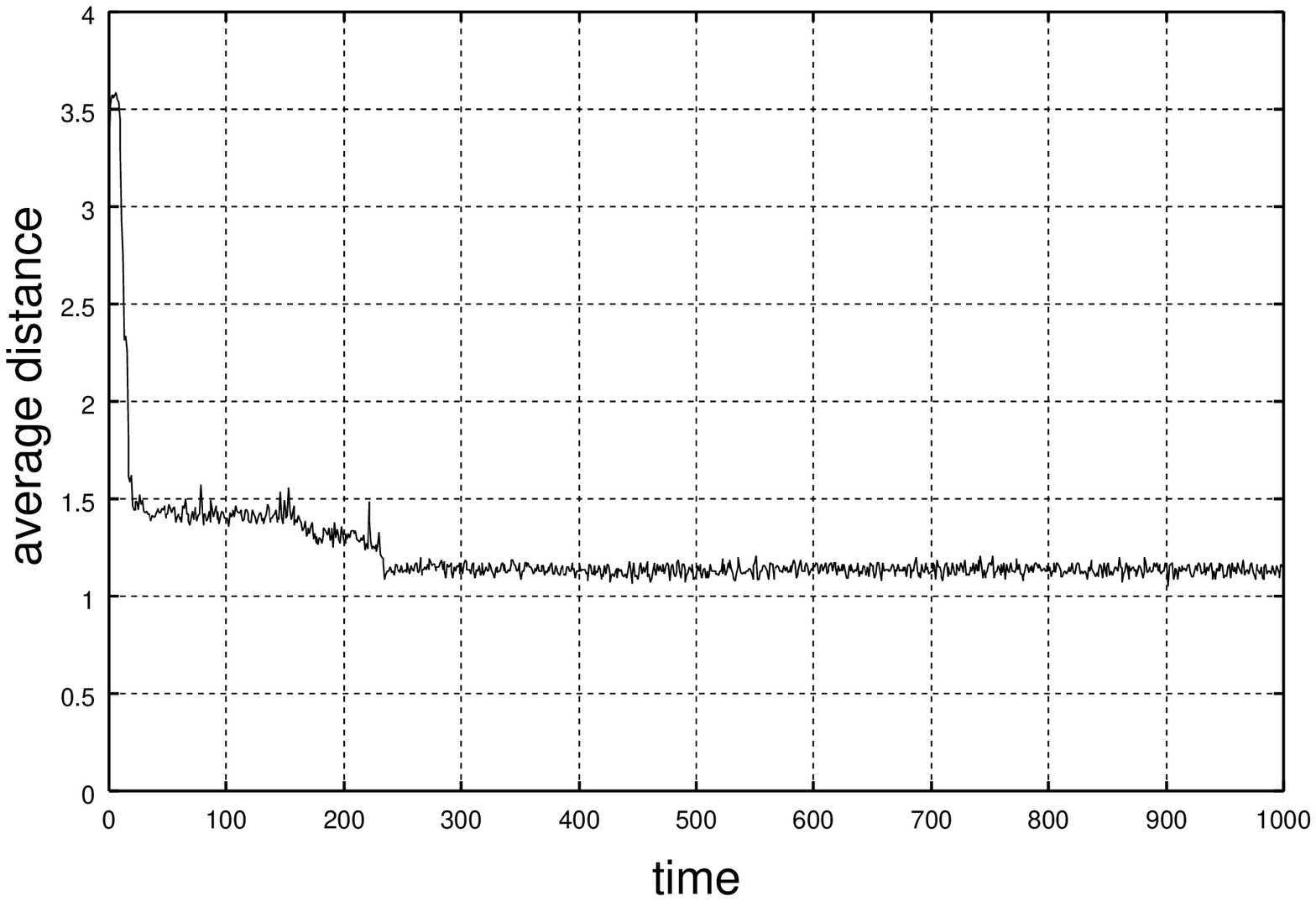}
\caption{
\small{The time series of the average attractor distance,
corresponding to the temporal evolution of Fig.\ref{fig14}.
Here the unit time of the figures is 1000.}
}
\label{fig15}
\end{center}
\end{figure}

We study how developmental process changes with the change of the noise 
amplitude. When the noise amplitude is too large, multi-cellular states 
change almost randomly in time. If the amplitude is small, stable
multi-cellular states of several cell types are formed eventually, and the 
switching process does not appear any more. An example of such long 
time behaviors is plotted in Fig.\ref{fig14}, and the corresponding change 
of the average attractor distance is also plotted in Fig.\ref{fig15}.
Starting from different initial conditions, different multi-cellular
states consisting of different cell types are realized, which have
different plasticity.  
It is now shown that multi-cellular states (with relatively
low plasticity) are stabilized by the noise.

\begin{figure}[ht]
\begin{center}
\includegraphics[width=4in,height=2in]{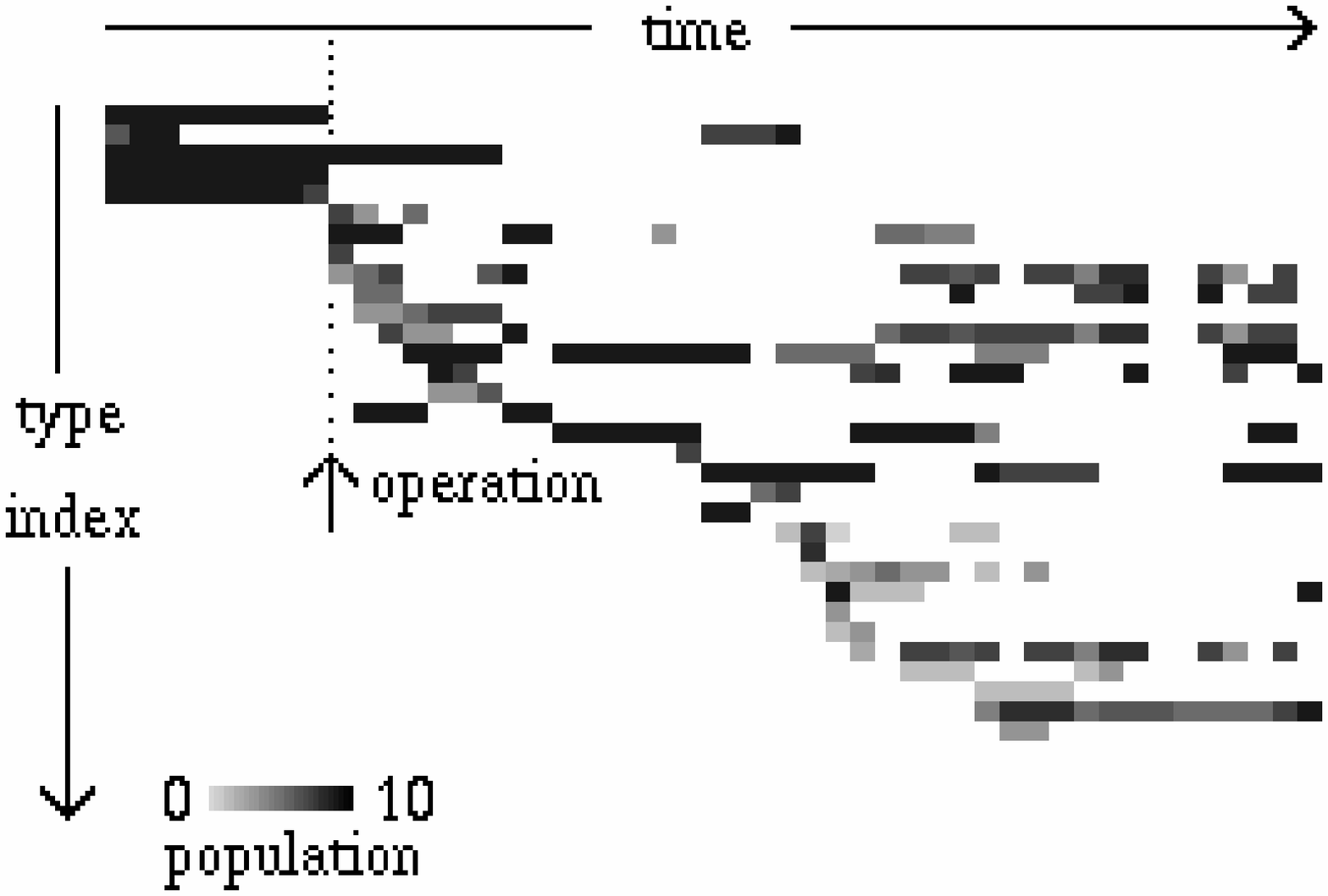}
\caption{
\small{An example of the temporal behavior of cell differentiation process
after external change of environmental resources shown by the arrow. 
Initially, cell distribution was given by the quasi-stable multi-cellular 
state at $t=10^6$ in the simulation of Fig.\ref{fig14}. Then, the simulation 
is carried out up to $t=50000$, by reducing the supply of five resource 
chemicals at $t=5000$, as shown by the arrow. 
The states of all cells are sampled by every $1000$ time unit, to classify 
all cell types. The method for the classification of cell types is mentioned 
in the text. Here, indices of cell types are numbered in the order of their
appearance. The population of each cell type at each time is represented 
with a gray scale.}
}
\label{fig16}
\includegraphics[width=4in,height=2in]{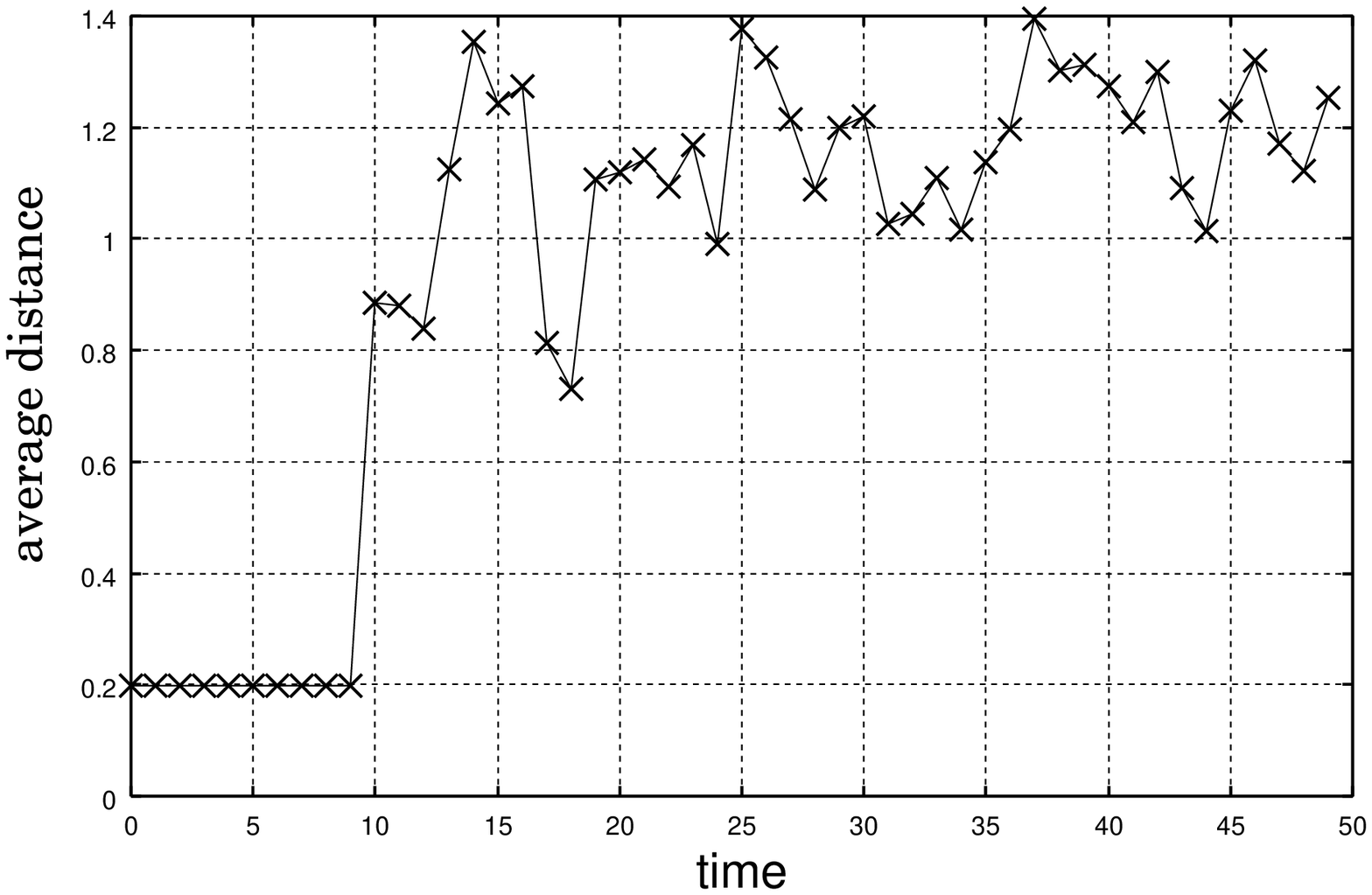}
\caption{
\small{The time series of the average attractor distance,
corresponding to the temporal evolution of Fig.\ref{fig16}.
Here the unit time of the figures is 1000.}
}
\label{fig17}
\end{center}
\end{figure}

Next we study the behavior against external change of the environment. 
As an example, we decreased the supply of some resource chemicals.  
The change of multi-cellular states as a result of the restriction of
some resources is shown in Fig.\ref{fig16}.  
When concentration of five chemical resources are reduced, given by the 
arrow in the figure, the original cell types that have low plasticity 
become unstable, and some of the cell types regain the plasticity.
Then the cell differentiation process is restarted,
leading to a novel multi-cellular state.
The corresponding change of the average attractor distance is also plotted 
in Fig.\ref{fig17}, which clearly 
shows that the plasticity is regained by the external change of environment.
Thus, a multi-cellular states that was stable and fixed is destabilized
by external operation, which leads to the change of environment.

\section{Summary and discussion}

In this paper, we have studied a dynamical systems model of developmental 
process, 
by introducing a new framework, namely, 
reaction-diffusion system on `chemical species space' for  
intra-cellular chemical reaction dynamics. 
By taking the developmental process into account further, 
it is shown that cells are differentiated 
into several types. The condition for the cell differentiation by cell-cell 
interaction is obtained.

As a long-term behavior, we have found the switching over several 
multi-cellular states that maintain diverse cell types.
In each multi-cellular state, diverse cell types coexist to reduce the 
competition for chemical resources, while the switching
is characterized by multiple cell deaths arising from the loss
of diversity of cell types and higher competition for the resources.
This switching behavior is first discovered in the present model.

\begin{figure}[b]
\begin{center}
\includegraphics[width=5in,height=4.5in]{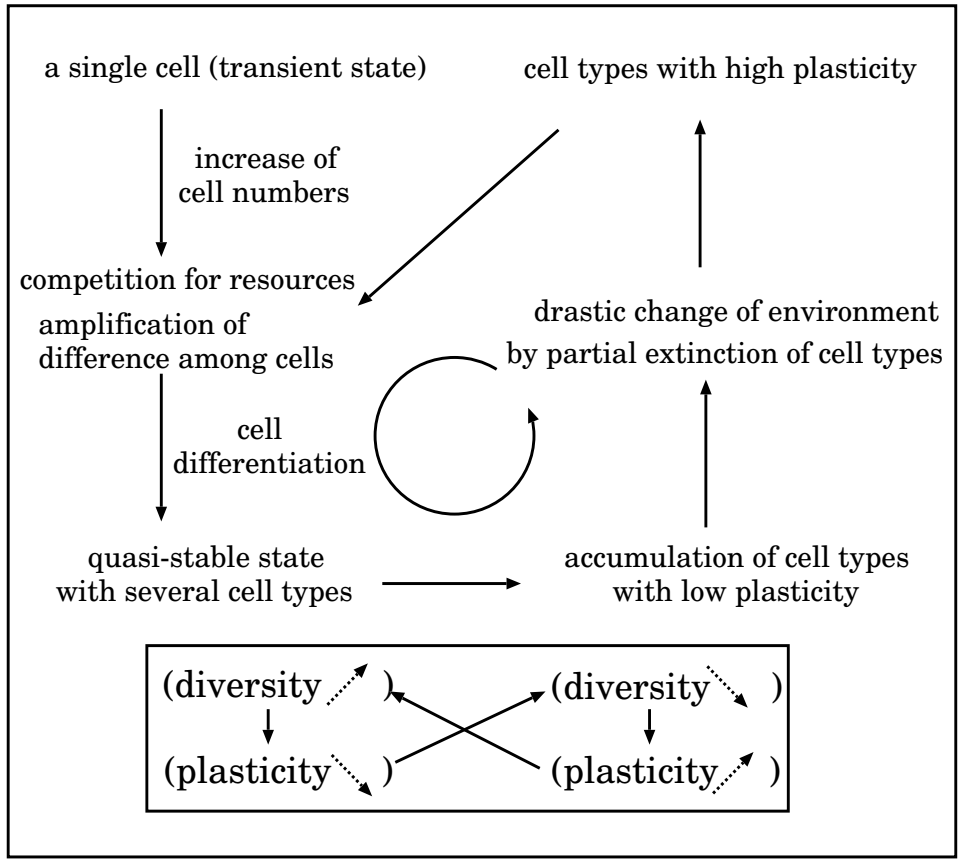}
\vskip 0.25cm
\caption{
\small{Schematic representation of our scenario for the mechanism
to keep the diversity of cell types.}}
\label{fig18}
\end{center}
\end{figure}

Then, we propose that this switching behavior is characterized by the
loss of plasticity of total cells, that is a general consequence of 
our dynamical systems theory.
The irreversible loss of plasticity is a 
general course in the developmental process, 
i.e., differentiation from a cell type 
with relatively high plasticity to that with lower plasticity, so that the 
ratio of cell types with lower plasticity increases gradually.
Then, effective use of resource chemicals by a suitable distribution of 
different cell types is destroyed, which leads to multiple cell deaths. 
Drastic change of the composition of environmental resource chemicals is 
resulted. Cells with low plasticity are replaced by those
with high plasticity.
As a result, a new multi-cellular state with novel cell types is generated. 
This process is repeated.
Schematic representation of the above scenario is summarized in 
Fig.\ref{fig18}.

The existence of several quasi-stable multi-cellular states
is important to consider the origin of several
tissues in multi-cellular organisms.
In multi-cellular organisms, several tissues coexist that are
represented as a cell ensemble with a different composition of 
consisting cell types, with a common gene set.
Then, the switching over several multi-cellular quasi-stable states will 
be important to study how the life cycle of a multi-cellular 
organism is formed. As a first step toward such study, we impose 
some external change to the system to make a transition between different 
multi-cellular states. 
In future, the search for a rule of transitions
between successive multi-cellular states will be important.
The dynamics of metamorphosis can be discussed along the line.

In our switching process, all cells with low plasticity are changed to those
with high plasticity, through multiple, simultaneous cell deaths.
In real metamorphosis, such kind of ``extinction" of many cells is also 
observed generally, as seen for example in insects.
According to our results, fluctuations of cell states have positive 
correlation with the plasticity of cell types. 

The present scenario of loss of plasticity and recovery by multiple
cell deaths can be experimentally verified, by
measuring the gene expression as indicators of chemical concentrations.
One can measure the variance of gene expressions, for example
by using flourscent protein and cell sorter,  during the course
of the developmental process.  Consider generally growth of a colony of
cells. By measuring the change of variance
of gene expressions over cells\cite{Elowitz,Collins,Kashiwagi} 
through the developmental process of cells, 
one can examine whether there is a decrease
in fluctuations, and moreover, whether there is a recovery 
when nultiple cell deaths lead to a novel ensemble of cells,
as in the event of metamorphosis.   
\\
In the present paper, we discussed spontaneous cell differentiation and 
switching processes in a well-stirred medium, i.e., in a spatially 
homogeneous medium.
To discuss morphogenesis with spatial pattern formation,
it will also be interesting to include spatial inhomogeneity
in the medium.

The present results also have some implications to the evolution.
Indeed, one can extend our model to regard each unit as an organism,
instead of a cell, and include genetic change (mutation) as that
of parameters in the model.  With this extension,
different types in our model can be regarded as
different species. Indeed, a theory of sympatric 
speciation with using phenotypic plasticity is
recently  proposed along this line\cite{Kaneko&Yomo4,Takagi1}.
In a preliminary study of the present model including
genetic mutation process, we have observed sympatric 
speciation process to form several species,
while with this process, the plasticity of each species defined in 
this paper decreases. 
With this extension, successive extinction 
events of some cell types in the present model correspond to
mass extinction of species through evolution.
Note that in the theory of punctuated equilibrium\cite{Gould}, evolution
process consists of long quasi-stationary regime and rapid temporal change
accompanied by extinctions, as discussed above.

Here, the recovery of plasticity 
of species after extinction of many cells is relevant to
open-ended evolution, as will be discussed elsewhere.

acknowledgments\\
We would like to thank T.Yomo and C.Furusawa for stimulating discussions. 
This work is supported by Grants-in-Aid for Scientific Research from
the Ministry of Education, Science and Culture of Japan
(11CE2006, Komaba Complex Systems Life Project).

\appendix
\section{The condition for cell differentiation}

Here we study conditions for cell differentiation.
First, we study the initial condition dependence of differentiation event.
As is mentioned above, it is necessary for an initial cellular state not to
be exactly at an attractor of a single cell. 
If we start precisely from an attractor of a single cell state, then the cells 
that are derived from its successive divisions cannot differentiate. 
Once the state is on an attractor, the cell division gives two identical cells,
as long as the fluctuation in cell division is not large to make
different attractor.
Then, with the increase of cell numbers, all the cells compete 
for the same resources, and eventually they come to the stage that
all the cells cannot take enough resources leading to the decrease
in the chemical contents in a cell.  Hence all the cells die,
almost simultaneously at some stage.
On the other hand,
cell differentiation, however, generally emerges as long as the initial
condition is not chosen precisely on an attractor of a single cell state.

Second, we study dependence of the differentiation frequency on $C_{v}/C_{u}$,
the ratio of the time scale of species-changing reaction among resource 
chemicals to the one among product chemicals.
We changed the parameter $C_{v}$ from 0.002 to 20 while fixing $C_{u}=2.0$.

If $C_{v}$ is much smaller than $C_{u}$, all the cells increase their number 
with keeping almost the same chemical composition, and competition for the 
resources is too strong for cells to survive.
Whereas, if $C_{v}$ is a comparable order of $C_{u}$, 
all the cells take the same uniform state with $u^j\ne1$ and $v^j\ne0$, 
so that extinction event occurs.
Cell differentiation events occur most probably in the intermediate case
between above two cases.

Third, we study the dependence of differentiation event on the strength
of interaction between cell and environment.
As an index of it, we change the parameter $D_{u}$ from 0.1 to 1.9. 
For $D_u \le 0.4$, cell-cell interaction is too weak to amplify
small differences among cell states, so that differentiation event cannot 
occur. Hence it is necessary that the strength of cell-cell interaction is 
stronger than some threshold value.

Fourth, we study the dependence of differentiation event on the number of
chemicals $K$. 
The frequency of cell differentiation rises with the increase of $K$. 
Hence it is necessary that the number of chemicals is larger than some value.\\
The necessary conditions for cell differentiation obtained in this appendix
are summarized in Sec.III.B:\\
(0) Initial condition is not set to be exactly on an attractor of a single 
cell.\\
(1) Inter-cellular interaction is stronger than some threshold.\\
(2) The ratio of the reaction time scale among $u^j$ to that among $v^j$ is 
in the intermediate range.\\
(3) The number of chemicals is beyond some level.

\clearpage

\begin{figure}[ht]
\begin{center}
\includegraphics[width=4in,height=2in]{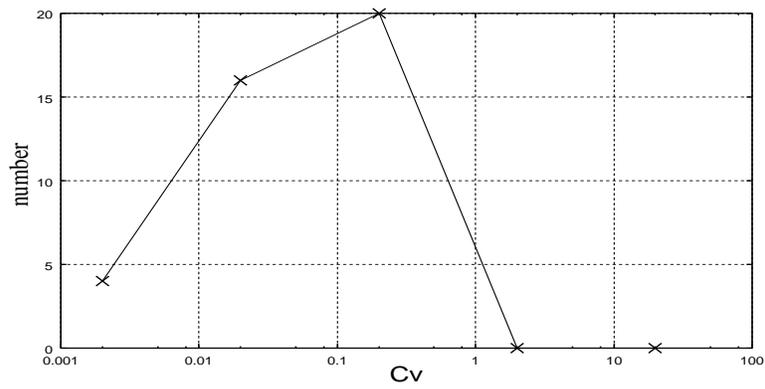}
\vskip 0.25cm
\caption{
\small{The frequency of cell differentiation plotted as a function
of the parameter $C_{v}$ with fixing $C_{u}=2.0$.
For each point of the figure, we took 20 different initial conditions,
and carried out simulations up to $t=20000$, to check the differentiation.
Plotted are the number of events with cell differentiation.}
}
\label{fig21}
\end{center}
\end{figure}

\begin{figure}[ht]
\begin{center}
\includegraphics[width=4in,height=2in]{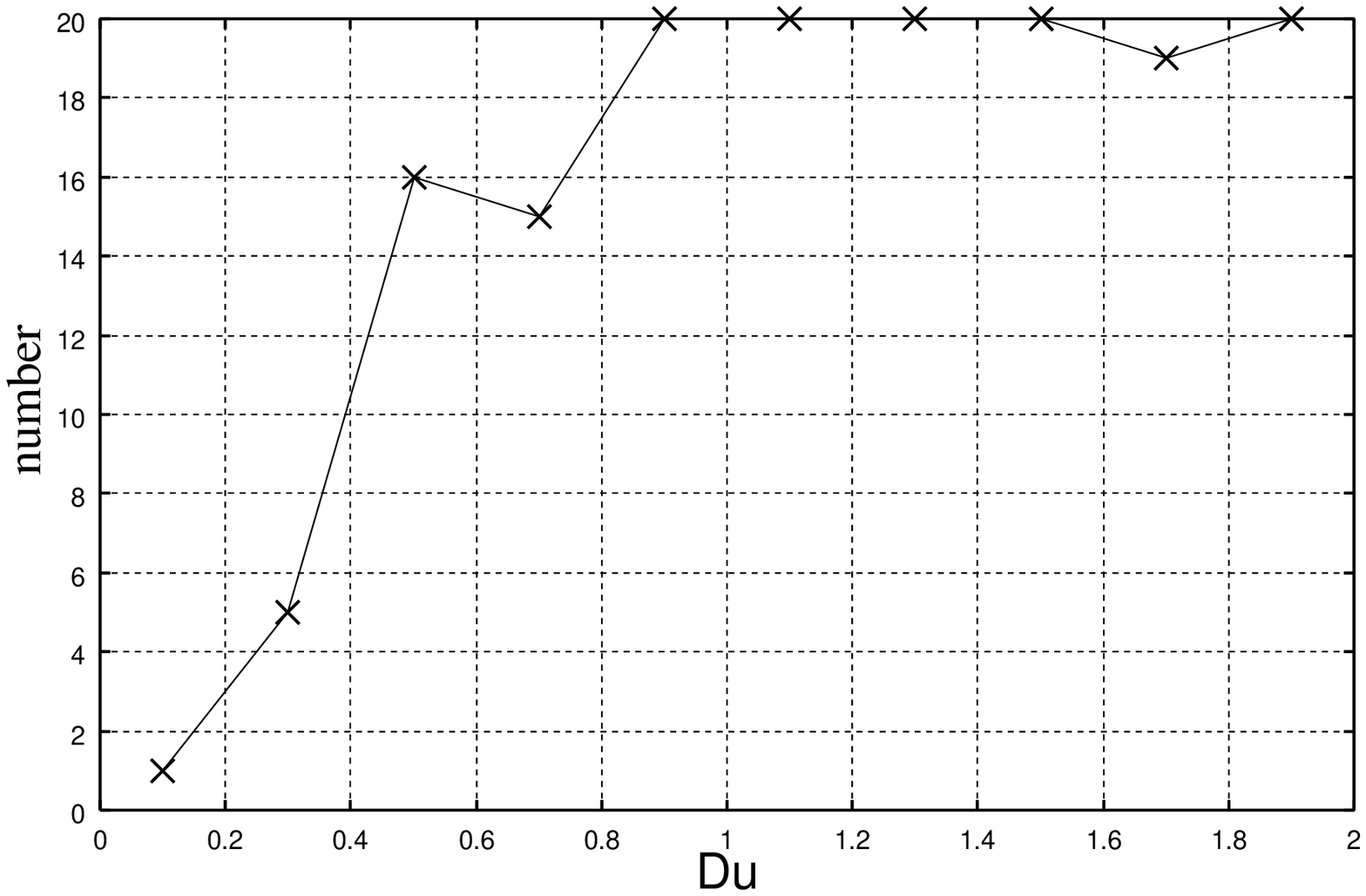}
\caption{
\small{The frequency of cell differentiation plotted as a function
of the parameter $D_{u}$. The parameter $D_{u}$ 
is changed from 0.1 to 1.9 with the bin size 0.2. 
For each point of the figure, we took 20 different initial conditions, 
and carried out simulations up to $t=20000$, to check the differentiation. 
Plotted are the number of events with cell differentiation.}
}
\label{fig22}
\includegraphics[width=4in,height=2in]{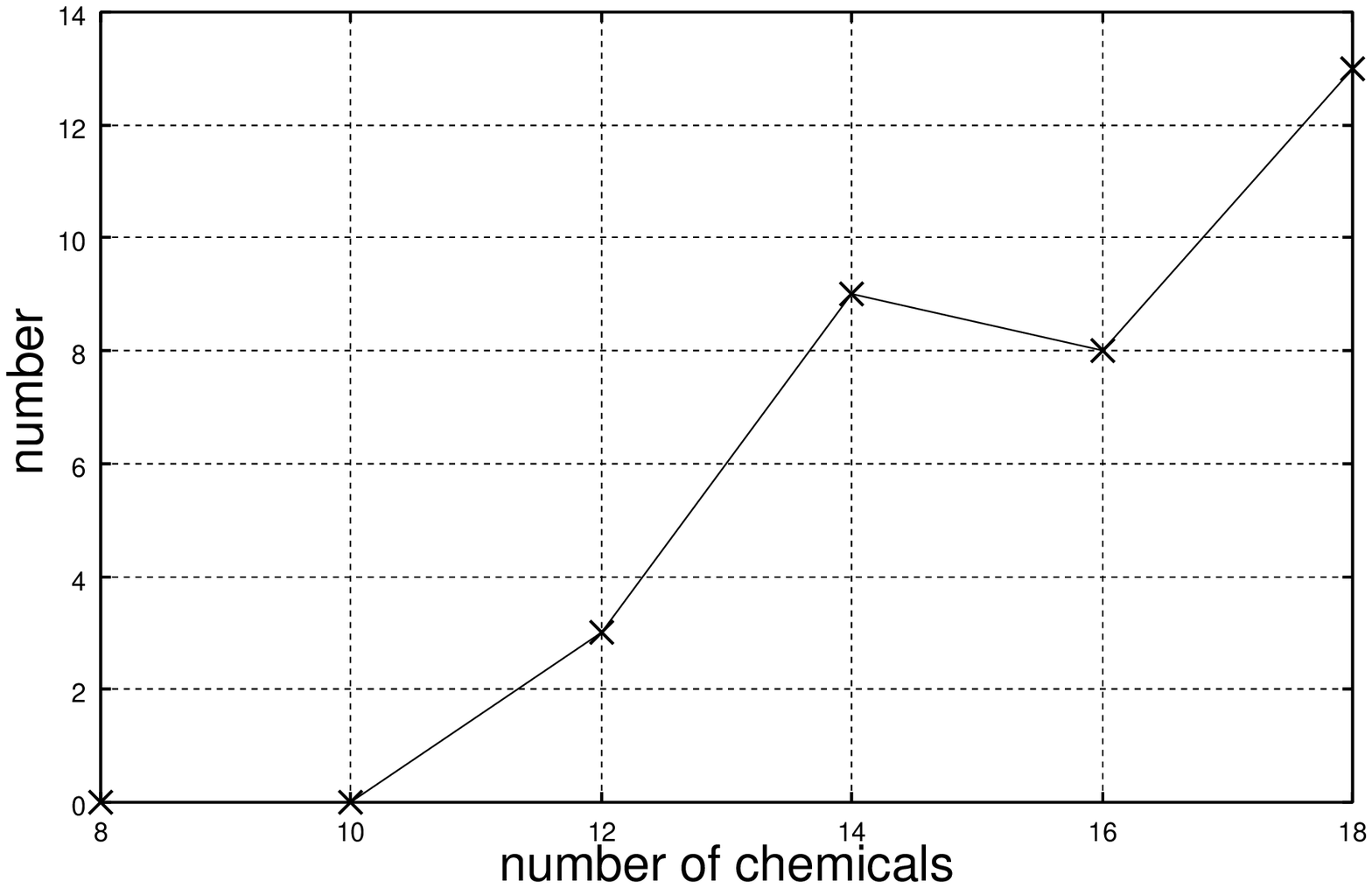}
\caption{
\small{
Dependence of cell differentiation on the number of chemicals $K$. 
We study the cases with $K=8$, 10, 12, 14, 16 and 18. 
For each point of the figure, we take 20 different initial conditions and
carry out a simulation up to $t=20000$ with the same parameters mentioned 
in the text. 
Accordingly we plot the number of the event where cell differentiation 
occurs.}
}
\label{fig23}
\end{center}
\end{figure}

\end{document}